%% file: samplepaper.tex
\documentclass[runningheads]{llncs}
\usepackage[T1]{fontenc}
\usepackage[utf8]{inputenc}
\usepackage{graphicx}
\usepackage{microtype}
\usepackage{cite}
\graphicspath{{img/}}
\usepackage[hyphens]{url}
\usepackage{hyperref}
\usepackage{graphicx}
\usepackage{amssymb}
\pdfoutput=1

\begin{document}
\title{Mechanisms and Attributes of Echo Chambers in Social Media}

%
%
\author{Bohan Jiang\orcidID{0000-0001-8552-2681} \and
Mansooreh Karami\orcidID{0000-0002-8168-8075} \and
Lu Cheng\orcidID{0000-0002-2503-2522} \and 
Tyler Black\orcidID{0000-0002-3526-8635} \and Huan Liu\orcidID{0000-0002-3264-7904}}
%
%
\institute{Computer Science and Engineering, Arizona State University, Tempe, AZ 85287, USA \\
\email{\{bjiang14, mkarami, lcheng35, tqblack, huan.liu\}@asu.edu}}
\maketitle              

\begin{abstract}
Echo chambers may exclude social media users from being exposed to other opinions, therefore, can cause rampant negative effects. Among abundant evidence are the 2016 and 2020 US presidential elections conspiracy theories and polarization, as well as the COVID-19 disinfodemic. To help better detect echo chambers and mitigate its negative effects, this paper explores the mechanisms and attributes of echo chambers in social media. In particular, we first illustrate four primary mechanisms related to three main factors: human psychology, social networks, and automatic systems. We then depict common attributes of echo chambers with a focus on the diffusion of misinformation, spreading of conspiracy theory, creation of social trends, political polarization, and emotional contagion of users. We illustrate each mechanism and attribute in a multi-perspective of sociology, psychology, and social computing with recent case studies. Our analysis suggest an emerging need to detect echo chambers and mitigate their negative effects.


\keywords{Echo chambers \and Misinformation \and Social media}
\end{abstract}
%
%

%
\section{Introduction} \label{intro}
The attributes of social media including low cost, easy access, and fast information dissemination mean that more people are consuming news from social media instead of traditional news outlets~\cite{shu2017fake}. However, both high-quality and low-quality information have been spread widely. Social media also limits the exposure to diverse opinions and forms groups of like-minded users~\cite{cinelli2021echo}. This results in echo chambers: a place in which people only encounter beliefs, opinions, or views that reflect and reinforce their own~\cite{jamieson2008echo}. Recent events such as the 2016 US presidential election and COVID-19 infodemic have evidence shown that trolls, shills, and cyborgs are actively peddling misinformation in social media~\cite{grinberg2019fake, gottlieb2020information, del2016spreading}. The recommender algorithms behind social media exacerbate the echo chambers by consistently presenting narratives about similar stances to users~\cite{pariser2011filter}. 

To contribute to the fight against this problem, we explore the individual and societal behaviors with existing literature. Four primary mechanisms underlying echo chambers effects are discussed. At their core, technology\let\thefootnote\relax\footnote{\scriptsize \textbf{Proceedings of the 2021 SBP-BRiMS} ,~Working/Late-Breaking Track,~July 6-9, 2021, Virtual} (e.g, the recommender algorithms of social media platforms) is the main cause of the proliferation of echo chambers in social media. Human psychology (e.g, confirmation bias and cognitive dissonance) and social networks(e.g, homophily) are also considered because they reinforce the existence of each other via a correlated feedback loop. To develop better echo chamber detection models, it is imperative to understand the mechanisms and common attributes beforehand. Previous studies describe echo chambers in different perspectives~\cite{barbera2015tweeting, bruns2017echo, jiang2021social}, but we focus on three crucial features to discuss the primary mechanisms and common attributes of echo chambers in social media: (1) Echo chambers are a network of users in social media; (2) the content shared in that network is one-sided and very similar in the stance and opinion on different topics; and (3) outside voices are discredited and actively excluded from the discussion. 




\section{Mechanisms of Echo Chambers} \label{mech}
\input{causes/cause}

\section{Attributes of Echo Chambers} \label{effect}
\input{effects/effects}

\section{Conclusions and Future Work}
\input{conclusion/conclusions}


\section*{Acknowledgements}
This material is based upon work supported by ONR (N00014-21-1-4002) and Parallax Advanced Research Corporation (11076-ASU).

%
%
%
\bibliographystyle{splncs04}
\bibliography{allref}
%

\end{document}

%% file: causes/cause.tex
\begin{figure}[hbt!]
\centering
\includegraphics[width=0.55\textwidth]{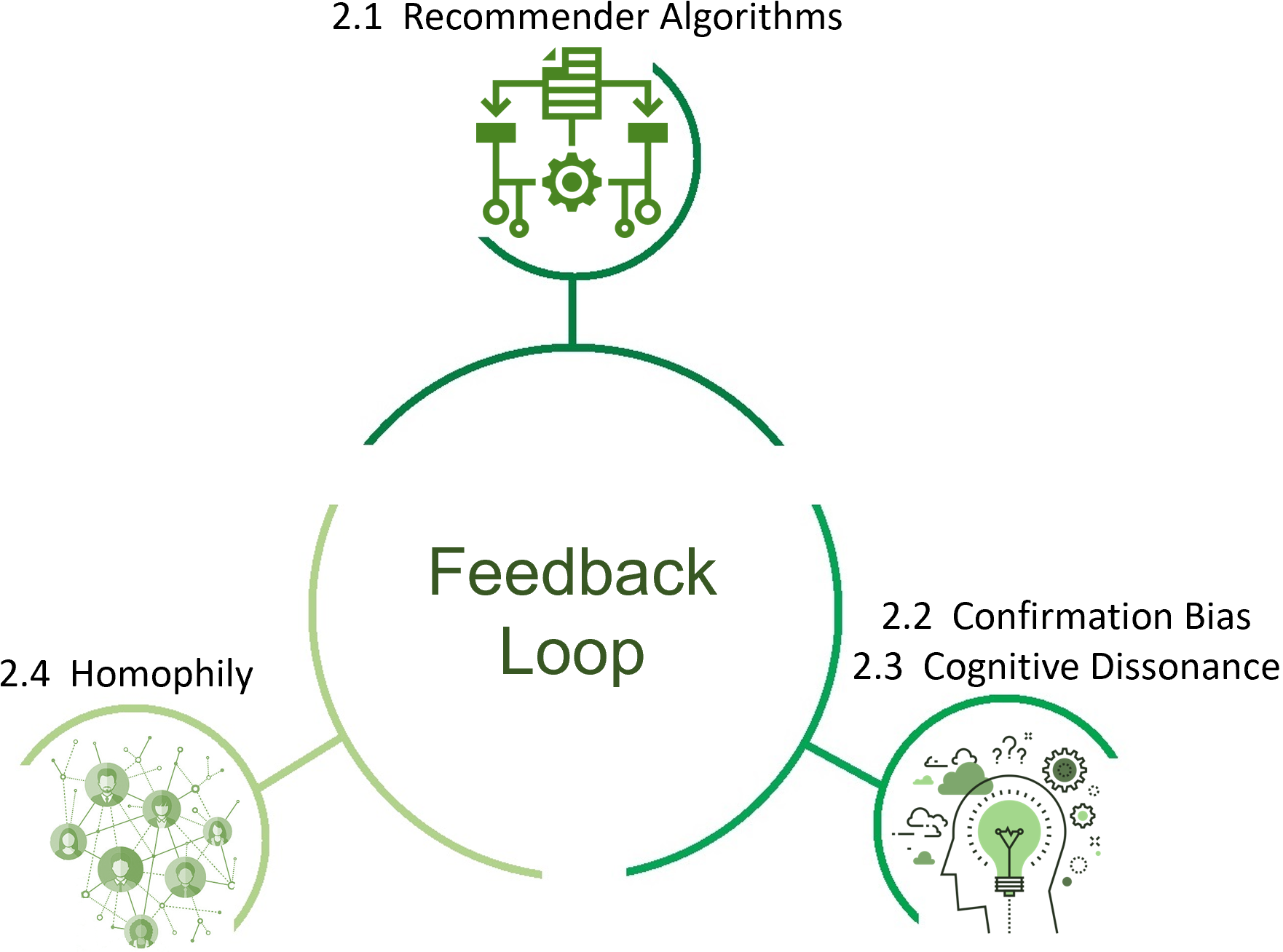}
\caption{Primary mechanisms underlying echo chamber effect related to three main factors: automatic systems such as recommender algorithms (Section~\ref{subsec::RecomAlgo}), human psychology such as confirmation bias and cognitive dissonance (Sections~\ref{subsec::ConfBias} and~\ref{subsec::CogDiss}), and social networks such as homophily (Section~\ref{subsec::Homophily}). These mechanisms are not mutually independent but highly correlated in a way that ultimately create feedback loops that further reinforce the existence of each other.}
\label{mechanism}
\end{figure}

In this section, we detail the primary mechanisms underlying echo chambers shown in Figure~\ref{mechanism}: recommender algorithms related to automatic systems, confirmation bias and cognitive dissonance related to human psychology, and homophily related to social networks. 

\subsection{Recommender Algorithms}\label{subsec::RecomAlgo}
Recommender algorithms trap users into personalized information by using their past behaviors to tailor recommendations to their preferences~\cite{Rastegarpanah2018}. These prediction engines ``constantly create and refine a theory of who you are and what you will do and want next''~\cite{pariser2011filter}, which then forms a unique universe of information around each of us. For example, when clicking on a news article, we show our interest in articles on this topic. The recommender algorithms take note of our behavior and will present more articles about similar topics in the future. As the process evolves, we are getting more and more personalized information, which ultimately leads us to: (1)~becoming the only person in the formed universe, (2)~not knowing how information is recommended, and (3)~unable to choose whether to enter this process~\cite{pariser2011filter}. This self-reinforcing pattern of narrow exposure and concentrated user interest caused by recommender algorithms is an important mechanism behind the echo chamber effect. 

Among the many outcomes of such recommender algorithms, e.g., narrower self-interest, overconfidence, decreased motivation to learn, the likely exacerbated polarization has the most negative impact. For this reason, many researchers have criticized recommender algorithms for the increase in societal polarization~\cite{Rastegarpanah2018,Hannak2017,Ge2020}. For example, Dandekar et al.~\cite{Dandekar2013} showed how many traditional recommender algorithms used on internet platforms can lead to polarization of user opinions in society. Therefore, an important line of research studies how to diversify the recommendation results, e.g.,~\cite{jiang2019degenerate}. 

%
\subsection{Confirmation Bias}\label{subsec::ConfBias}
Confirmation bias is the tendency to seek, interpret, favor, and recall information adhering to preexisting opinions~\cite{nickerson1998confirmation}. According to the selective exposure research~\cite{frey1986recent}, we tend to seek supporting information while avoiding challenging information.
Echo chambers are among one of the many outcomes of confirmation bias. The rampant use of social media further amplifies the effect of confirmation bias on echo chambers. There are three types of confirmation bias: biased search for information~\cite{mynatt1978consequences}, biased interpretation of information~\cite{lord1979biased}, and biased memory recall of information~\cite{hastie1986relationship}. In the context of social media, for example, users not only actively seek news that is consistent with their current hypothesis but also interpret information in their own ways. Even if both the collection and interpretation are neutral, they probably remember information selectively to reinforce their expectations, i.e., \textit{selective recall effect}~\cite{hastie1986relationship}.  

Confirmation bias and the provision of recommender algorithms create a self-reinforcing spiral. As described in Figure~\ref{mechanism}, on one hand, recommender algorithms provide users with more of the same content based on their past behaviors to shape the future preference; on the other hand, users accept and even actively seek such information due to confirmation bias. The feedback loop between recommender algorithms and human psychology eventually leads to an echo chamber that shifts users' world view.
\subsection{Cognitive Dissonance}\label{subsec::CogDiss}
In the field of social psychology, cognitive dissonance refers to an internal contradiction between two opinions, beliefs, or items of knowledge~\cite{Festinger1957ADissonance}. For example, if someone eats meat but at the same time cares about the animals' life~\cite{Loughnan2014TheAnimals}. On the grounds that people strive towards consistency, they psychologically feel the pressure to reduce or eliminate the distress caused by dissonance. Festinger~\cite{Festinger1957ADissonance} introduced three major strategies for dissonance reduction: (1)~change one or more of the beliefs, opinions, or behaviors, (2)~increase consonance by acquiring new information or belief, (3)~forget or reduce the importance of the cognitions. Echo chambers are considered as one of the practices in reducing dissonance. People try to seek out for ideologically consonant platforms and interactions to avoid contact with individuals that confront their ideas~\cite{Evans2018OpinionChambers}. Moreover, ideological homogeneity in online echo chambers can encourage extremism. There are two aspects for this stimulation: (1)~one's commitment to their thought will increase dramatically if it has been written down and disseminated to a public audience~\cite{Cialdini2007Influence:Persuasion}. For example, the act of tweeting or posting contents on social media websites; and (2)~discussion with like-minded individuals as well as the social support will reinforce the correctness of that belief~\cite{frey1986recent}. For instance, liking tweets/posts and retweeting/reposting thus boosting attitude extremity~\cite{Bright2020EchoViews}. All of which are in support of decreasing individuals' cognitive dissonance.




\subsection{Homophily}\label{subsec::Homophily}
Homophily, also known as love of the same, is the process by which similar individuals become friends or connected due to their high similarity~\cite{Zafarani2014SocialIntroduction}. This similarity can be of two types: (1)~status homophily, and (2)~value homophily~\cite{McPherson2001BirdsNetworks}. Status homophily deals with people who connect due to similar ascribed (sex, race, or ethnicity) or acquired characteristics (education or religion). Value homophily involves grouping similar people based on their values, attitudes, and beliefs regardless of their differences in status characteristics~\cite{McPherson2001BirdsNetworks}. Depending on the echo chamber's ideology, the echo chamber can be formed due to status homophily, value homophily, or both.
Social media and other online technologies have loosened the basic sources of homophily such as geography and allowed users to bind homophilous relationships on other dimensions like race, ethnicity, sex, gender, and religion.
Moreover, homophily has predictive and analytic power on social media and can be measured by how the assortativity, also known as social similarity, of the network has changed over time and modeled using independent cascade
models (ICM)~\cite{Zafarani2014SocialIntroduction}. By measuring political homophily on Twitter, we can investigate whether the structure of communication is an echo chamber- or public sphere-like~\cite{colleoni2014echo}, or whether there is a homophilous difference between the echo chambers of conservative individuals and liberal ones~\cite{Boutyline2017TheNetworks}.

%% file: effects/effects.tex

In this section, we illustrate five common attributes of echo chambers: diffusion of misinformation, spreading of conspiracy theory, creation of social trends, political polarization, and emotional contagion of users. We also discuss their different outcomes, social impacts, and potential risks.

\subsection{Diffusion of Misinformation}
Nowadays, mainstream social media platforms are used by people due to easy access, low cost, and fast dissemination of news pieces~\cite{shu2017fake}. However, the quality and credibility of the content spread in social media is considered lower than traditional news media because of a lack of regulatory authority. Thus, people manipulate the public by leveraging echo chambers to propagate misinformation~\cite{tornberg2018echo}. Echo chambers exclude dissenting opinions, make users insist on their confirmation bias, and let misinformation go viral. 

Despite early efforts have been undertaken to mitigate online misinformation~\cite{shu2020combating, gottlieb2020information, shu2019detecting}, the COVID-19 related misinformation were widely spreading on social media as a global crisis. Existing methods have been ineffective for the COVID-19 infodemic because: (1)~the contents are novel and highly deceptive; (2)~the dissemination is rapid; and (3)~they require experts with domain knowledge to fact-check. Echo chambers in social media manipulate not only influencers but also common people to become misinformation spreaders. They enable users to intentionally or unintentionally disseminate misinformation faster~\cite{tornberg2018echo}. Misinformation spread in echo chambers usually contains three characteristics: (1)~similar misinformation is frequently scrolled and repeated to the users; (2)~the contents are inflammatory and emotional; and (3)~meant to mislead people by exploiting social cognition and cognitive biases. Because the diffusion of misinformation is one of the most common attributes of echo chambers in social media, echo chamber detecting methods should take it into consideration.


\subsection{Spreading of Conspiracy Theories}
Echo chambers in social media have provided fertile grounds for conspiracy theories to spread quicker. Existing research illustrates that various conspiracy theories have been circulating through mainstream media~\cite{grimes2020health, juhasz2017political}. Conspiracy theories are attempts to explain the ultimate causes of significant social and political events and circumstances~\cite{douglas2019understanding}.
Conspiracy believers use social media to find each other, disseminate conspiracy contents, and share fringe viewpoints. Conspiracy theories express and amplify anxieties and fears about losing control of religious, political, or social order~\cite{marwick2017media}. Unlike misinformation, conspiracy theories are often strongly believed by governments. This results in catastrophic impact to society. For example, AIDS denial by the government of South Africa, was estimated to have resulted in the deaths of 333,000 people~\cite{simelela2015political}. 

Despite the fact that there was no clear evidence about ``the 2020 US presidential election was stolen by Democrats'', Trump and his rallies kept tweeting this claim to supporters. As a result, Trump's supporters including Proud Boy, QAnon, and the Oath Keepers supporters participated in the deadly capitol riot~\cite{dave2021political}. Prior intelligence indicated that several organizations issued warning about this event and potential violence in the days leading up to the storming. Advance Democracy, Inc., an independent, nonpartisan, non-profit organization, reported 1,480 posts on Twitter from QAnon-related accounts about the scheduled capitol riot contained ``term of violence''. They also identified TikTok videos with more than 279,000 views that called for rebellion beforehand\footnote{\url{https://thehill.com/policy/technology/533450-trump-supporters-organized-the-capitol-riot-online}}. While several social media companies suspended the capitol riot-related posts and accounts from their platforms afterward, other conspiracy theories are shared and spread through social media as an attribute of echo chamber~\cite{shahsavari2020conspiracy}.

\subsection{Creation of Social Trends}
Social media presents temporal popular topics as social trends on the main website to attract user’s attention. Many studies have tried to discover the important factors that cause trending topics~\cite{mathioudakis2010twittermonitor, romero2011influence, wu2007novelty}. Asur et al.~\cite{asur2011trends} found that the resonance of the content with the users of the social network is crucial. They further define the measurement of ``resonance’’ in three parts: (1)~the novelty of content; (2)~the influence of members of the diffusion network; and (3)~the impact of media outlets when the topics originate in standard news media. Information with high ``novelty'', ``influence'', and ``impact'' can capture huge attention in a short time. Thus, information spread from echo chambers in social media have the capability to create trending topics due to their large scope, like-minded stance, and social influence. Despite social trends containing misinformed statements and false claims, they were presented by social media and reported by mainstream news media~\cite{marwick2017media}. Social media platforms, influencers, and news media outlets should take responsibility to carefully display and report social trends. Moreover, social trends can be supervised to detect echo chambers for malicious activities.

\subsection{Political Polarization}
\begin{figure}[hbt!]
\centering
\includegraphics[width=0.8\textwidth]{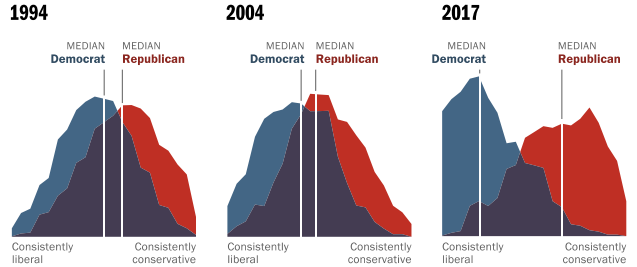}
\caption{Comparison of public political polarization in the U.S over the past two decades, seven Pew Research Centers collected surveys with 10 questions since 1994. Source from Pew Research Center, Washington, D.C. October 20, 2017.}
\label{polarization}
\end{figure}

In the 2020 US presidential election, both of the two main candidates received more than 74 million votes. Joe Biden created a historic record with a total of more than 81 million ballots received. It is clear that the United States experienced record levels of voter engagement, but it also means the country is extremely polarized. As shown in Figure~\ref{polarization}, the gap between two major parties in the US has increased while the overlapping has decreased significantly over the past two decades.
In social media, we can observe two giant partisan echo chambers represent two major political groups of people with opposite political opinion and stances~\cite{colleoni2014echo}. Given that individuals tend to align with those who are like-minded in nature, politicians and parties intentionally reinforce the partisan bias inside echo chambers, leading to an increasing level of political polarization~\cite{levy2019echo}. For example, Levy et al.~\cite{levy2019echo} illustrated that politicians made decisions and policies motivated by political purposes rather than social benefits. Political polarization can cause extreme selective exposure, cognitive bias, and correlation neglect~\cite{sears1967selective}. However, Dubois et al.~\cite{dubois2018echo} found that there is only a small segment of the population are likely to find themselves in an echo chamber. Essentially, the impact of partisan echo chambers is overstated. They suggested that echo chamber researchers should test the theory in the realistic context of a multiple media environment.  

\subsection{Emotional Contagion of Users}
Emotional states can be transferred to others via emotional contagion, leading people to suffer from the same emotions without their awareness~\cite{fowler2008dynamic, rosenquist2011social}. A recent study showed that extreme emotions are exposed and amplified by echo chambers~\cite{wollebaek2019anger}. This manifestation is usually caused by users who continually receiving misleading contents and conspiracy theories. For example, from a COVID-19 case study of China, Ahmed et al.~\cite{ahmed2020epidemic} illustrated that young people, aged 21-40 years old, were suffering from psychological problems during the COVID-19 epidemic. This is because young people who frequently participate in social media repeatedly receive broadcasts of fatality rate, confirmed cases, and misleading information via echo chambers. Moreover, Del et al. found that inside the echo chamber, active users appear to become highly emotional relative to less active users~\cite{del2016echo}. Their analysis indicated that the higher involvement in the echo chamber enables more negative mental behaviors. Kramer et al.~\cite{kramer2014experimental} provided experimental evidence that emotional contagion can occur without direct interaction between people, and in the complete absence of nonverbal cues. These types of echo chambers are difficult to detect in social media via content-based or network-based methods. 


%% file: conclusion/conclusions.tex
In times of crisis, whether political or health-related, misinformation, conspiracy theory, and extreme speech are widely spread and amplified by echo chambers in social media platforms. In this paper, we explored echo chambers in social media by reviewing existing literature in the fields of sociology, psychology, and social computing. In the mechanisms section, we discussed the key mechanisms that lead to the formation and growth of echo chambers. We illustrated that echo chambers were not only caused by the recommender algorithms, but also the specific confirmation bias, cognitive dissonance, and homophily among users. Next, by reviewing the external circumstances of recent events, we illustrated five common attributes. Our analysis indicated that the diffusion of misinformation is the most common attribute. Meanwhile, the spreading of conspiracy theory, creation of social trending, political polarization, and emotional contagion of users are represented as well. Further, the effects of leaderships, social media influencers, and news outlets cannot be diminished. Existing echo chamber detection models can only tackle this problem in a algorithmic aspect. More educational works should be encouraged for social media participants to understand the mechanisms, attributes, and risks of echo chambers. It is also promising to consider the role of sociological and psychological theories in recommender algorithms. 